# High-quality femtosecond laser surface micro/nano-structuring assisted by a thin frost layer

*Wenhai Gao#, Kai Zheng#, Yang Liao\*, Henglei Du, Chengpu Liu, Chengrun Ye, Ke Liu, Shaoming Xie, Cong Chen, Junchi Chen, Yujie Peng, and Yuxin Leng\**

W. Gao, K. Zheng, Dr. Y. Liao, H. Du, Dr. C. Liu, C. Ye, K. Liu, S. Xie, C. Chen, Dr. J. Chen, Dr. Y. Peng, and Dr. Y. Leng
Key Laboratory of High Field Laser Physics and CAS Center for Excellence in Ultra-Intense Laser Science, Shanghai Institute of Optics and Fine Mechanics, Chinese Academy of Sciences, Shanghai 201800, China
E-mail: superliao@siom.ac.cn, lengyuxin@mail.siom.ac.cn

K. Zheng
School of Physics and Optoelectronic Engineering, Hangzhou Institute for Advanced Study, University of Chinese Academy of Sciences, Hangzhou 310024, China

C. Chen
School of Optical-Electrical and Computer Engineering, University of Shanghai for Science and Technology, Shanghai 200093, China

Keywords: ((femtosecond laser, surface micro/nanostructures, frost-assisted, high-quality, high-spatial-frequency, laser induced periodic surface structures))

Femtosecond laser ablation has been demonstrated to be a versatile tool to produce micro/nanoscale features with high precision and accuracy. However, the use of high laser fluence to increase the ablation efficiency usually results in unwanted effects, such as redeposition of debris, formation of recast layer and heat-affected zone in or around the ablation craters. Here we circumvent this limitation by exploiting a thin frost layer with a thickness of tens of microns, which can be directly formed by the condensation of water vapor from the air onto the exposed surface whose temperature is below the freezing point. When femtosecond laser beam is focused onto the target surface covered with a thin frost layer, only the local frost layer around the laser-irradiated spot melts into water, helping to boost ablation efficiency, suppress the recast layer and reduce the heat-affect zone, while the remaining frost layer can prevent ablation debris from adhering to the target surface. By this frost-assisted strategy, high-quality surface micro/nano-structures are successfully achieved on both plane and curved surfaces at high laser fluences, and the mechanism behind the formation of high-spatial-frequency (HSF) laser induced periodic surface structures (LIPSSs) on silicon is discussed.

## 1. Introduction

Ultrafast laser surface micro/nano-structuring has emerged as an enabling technology with a wide variety of applications involving self-cleaning, anti-reflection, anti-microbe and anti-friction, due to its high flexibility in tuning optical, mechanical, or chemical properties at surfaces and interfaces of materials. [1] In the meantime, recent rapid advances in ultrafast laser technology have made high-average-power ultrafast lasers increasingly reliable and cost-effective for high-throughput material processing.[2,3] Although ultrafast laser ablation, under the proper control of parameters, has proved its uniqueness to produce micro/nanoscale features with high precision and accuracy,[4] there still remain several inevitable issues as the laser fluence increased, such as redeposition of debris, formation of recast layer and heat-affected zone in or around the ablation regions. The debris and recast layer, which was formed by rapid cooling and re-solidification of high-temperature plasma or molten material generated by laser ablation, usually has strong adhesiveness and is difficult to be removed.[5,6] Meanwhile, located on the surface close to the irradiated zone or in the ablated craters and grooves, the debris and recast layer may block the propagating path of subsequent laser pulses. In this regard, the ablated debris and recast layer not only affect the surface quality and deteriorates the functionality, but also reduce the ablation efficiency.

Many efforts have been made to remove the debris and recast layer in the laser ablation process.[7-12] Conventionally, water-assisted laser ablation has been extensively exploited to remove the debris, in which the debris can be carried away by thermal convection of liquid and bubble-induced liquid motion. Nevertheless, water-assisted laser processing suffers from scattering or deviation of the incident laser by the water surface, cavitation bubbles and suspended ablated material.[12] An ice shield layer recently was introduced to prevent debris redeposition while reducing surface disturbance,[13,14] in which the ice layer was preliminarily formed by freezing a water layer on the sample surface, and locally melted or vaporized in the process of laser ablation. However, the ice layer obtained by freezing water usually has a thickness more than several hundred microns due to the large surface tension of water. For femtosecond laser precise ablation, it is difficult to melt or remove the thick ice layer by tightly focused femtosecond pulses with low pulse energy.

Here, we present a novel approach towards high-quality surface micro/nano-structuring by femtosecond laser ablation assisted by a thin layer of condensed water frost on cold surfaces. By controlling sample surface temperature and relative humidity of the surrounding atmosphere, a thin frost layer with a thickness of tens microns could be homogeneously deposited on all surfaces of the sample. When a femtosecond laser was focused on the sample surface covered with the thin frost layer, the frost located in the focusing area would rapidly melted to a thin layer of water, helping to disperse debris and reduce recast layer in subsequent laser ablation, while the remaining frost layer could prevent ablation debris from adhering to the target surface. The ablation efficiency and surface micromorphology were comparatively studied by femtosecond laser ablation on different materials with and without the assistance of frost layer. High-quality micro/nanostructures on both plane and curved surfaces of different materials were demonstrated by the frost-assisted strategy, and the mechanism behind the high-quality and high-precision nanostructures was discussed.

## 2. Experimental procedures

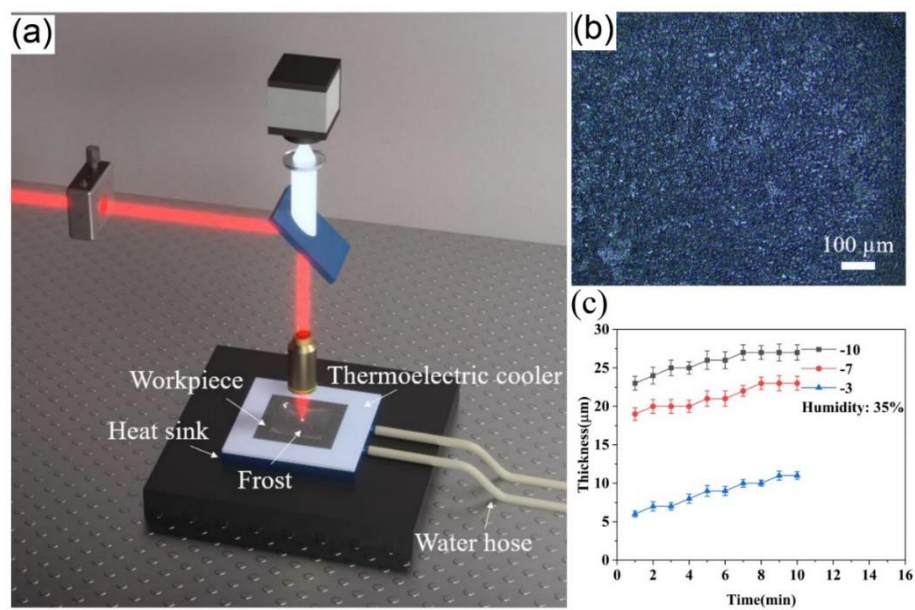

**Figure 1.** (a) Schematic of frost-assisted femtosecond laser ablation on the flat surface; (b) Optical microscopic image of thin frost layer formed on silicon plate; (c) Frost layer thickness as a function of condensation time at different cold surface temperatures, in an indoor air environment with a relative humidity of 35%.

The schematic of our experimental setup is illustrated in **Figure 1a**, which mainly consists of a femtosecond laser source, an objective lens mounted on a Z-axis translation stage, a CCD camera system for optical microscopy and a thermoelectric cooling system fixed on an XY translation stage. To perform precision surface structuring, an industrial femtosecond fiber laser was used to produce linearly polarized laser pulses with a center wavelength at 1030 nm, a pulse duration of ~500 fs and a repetition rate of 100 kHz. Standard single-crystalline silicon wafers ((100)-oriented, single side polished, 500-μm thick), fused silica glass pieces and borosilicate glass rods (Schott Borofloat 33, 10-mm diameter) were used as the workpiece materials. Several objective lenses with different numerical apertures (N.A.) were employed for focusing the laser beam onto the workpieces. To form the frost layer, a thermoelectric cooler sandwiched between the workpiece and a heat sink was used to keep surface temperature of the workpiece below the freezing point. The surface temperature was automatically controlled by a closed-loop temperature control system including a temperature detector and a proportional integral differential (PID) controller.

The thickness of frost layer on the workpiece was measured by a three-axis measuring microscope with a 50× objective lens (N.A. = 0.85) and an axial resolution of ~1 μm. The surface morphology of ablation patterns was obtained by an optical microscope (Olympus BX43) and a field emission scanning electron microscope (SEM, Tescan Mira3). The 2D depth profiles of the ablated craters were captured by use of a white-light interferometric microscope (Bruker Contour GT). Before microscope observations, all the samples were rinsed with distilled water for 5 min and dried at room temperature. The laser power and pulses energy were measured on the target surfaces with a pyroelectric detector (Thorlabs, S442C).

## 3. Results and Discussion
### 3.1. Formation of a thin frost layer

Frosting is a ubiquitous natural phenomenon by condensing water vapor in a humid environment onto the exposed surface whose temperature is below the freezing point. During frost formation process, the water vapor in the air can continuously freeze on and between the existing frost crystals, which contributes to the increase in thickness and density of frost layer.[15] Figure 1b presents a microscope image of the thin frost layer on the silicon wafer, in which the cooled silicon wafer was exposed to ambient air and its surface temperature was maintained at -5°C. The ambient temperature and relative humidity were kept constant at 25°C and 35%, respectively. It can be observed that the frost layer was relatively uniform and dense over the whole surface, and the average size of the frost crystals was about 4 μm. The dense frost crystals can be attributed to its relatively high frost surface temperature close to the melting temperature,[16] which is helpful for achieving a good consistency of frost-assisted laser processing.

Equation S1, Supporting Information, suggests that the slow rate of frost growth depends on high frost surface temperature, low ambient humidity, and low convective mass transfer coefficient. In our experiments, to obtain a thin frost layer with stable thickness, the surface of silicon wafer was maintained at a temperature slightly below the freezing point, and the relative humidity of ambient air was set at 35%. Figure 1c presents the evolution of frost thickness over condensation time at different cold surface temperatures, showing that the thickness of frost increases slowly with time. It can be seen that a frost layer with a thickness in the range of 5-15 μm can be obtained in a condensation process of 10 min. The stability of frost layer thickness could be further improved by using an on-line thickness measurement system in future.

## 3.2. Surface morphology and ablation rate

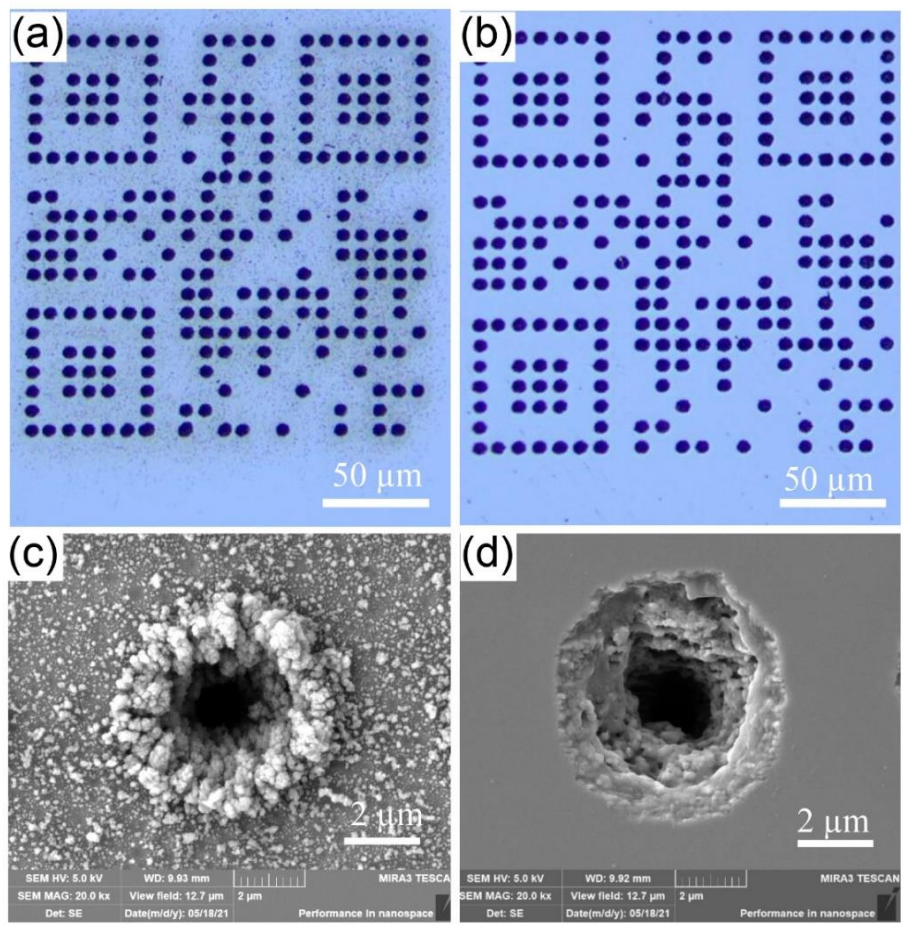

**Figure 2.** Optical micrographs of two-dimensional barcodes inscribed on silicon covered without (a) and with (b) frost layer; SEM micrographs of a single ablated crater inscribed on silicon covered without (c) and with (d) frost layer.

**Figure 2**a and 2b show the micromorphology of two-dimensional (2D) barcodes inscribed on a silicon plate without and with frost assistance, respectively. The barcodes consist of an array of ablated craters ablated by tightly focused femtosecond laser (N.A.=0.45). The single ablated crater was produced by 1000-pulse irradiation at a pulse energy of 2.5 µJ. One can see that there is a large amount of debris around the ablated craters, and thick recast layer is formed in the craters for the traditional processing. In contrast, for the frost-assisted counterpart, the silicon surface is very clean and no any resolidified material can be seen in or around the craters. This phenomenon may be attributed to rapid melting of the thin frost layer under irradiation of femtosecond pulses. The ice-water phase transition after an ultrafast temperature jump could occur on a time scale of 100 ps.[17] During this period, heat conduction to the silicon substrate and the lateral frost layer is negligible.[18] Therefore, a local and thin water film could be instantaneously produced under the irradiation of first laser pulses, resulting in clean and debris-free ablated craters by successive laser pulses. These results are consistent with previously reported femtosecond ablation under sprayed thin water film.[19] Nevertheless, the thickness of frost layer is more uniform and controllable over the sample surfaces in comparison with water film.

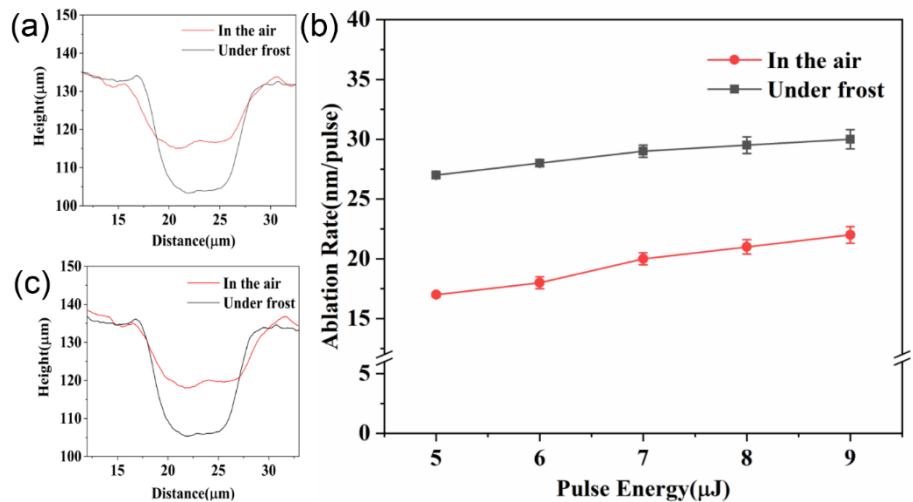

**Figure 3.** Crater depth profiles in the air and under thin frost layer after 1000-pulse irradiation at different pulse energies: (a) 5 µJ and (b) 9 µJ; (c)Ablation rates of silicon as a function of laser pulse energy in the air and under thin frost layer.

The efficiency of frost-assisted laser ablation was evaluated by comparing profiles of craters. **Figure 3**a and 3b present 2D depth profiles of the ablation craters after multiple-pulse irradiation on silicon covered without and with frost layer. The thickness of frost layer was controlled at ~ 20 µm and a 10× objective lens (N.A. = 0.3) was used to produce ablation craters with a diameter of ~ 12 μm. One can see that the crater depths by frost-assisted ablation are significantly larger than those without frost layer, while the crater diameters remain roughly unchanged. For different pulse energies from 5 μJ to 9 μJ, an approximately 40~60% improvement in ablation rate was found, as shown in Figure 3c. Here, the ablation rate is defined as the average crater depth divided by the number of applied laser pulses. This improvement could be based on the same mechanism for water-assisted laser ablation[20]: Water from melting frost disperses the ablation debris and reduces the recast layer, resulting in less blockage for successive laser pulses into the ablation craters. Note that the 40~60% improvement of ablation rate by frost-assisted ablation is lower than the previously reported results achieved by use of a thin layer of water film with thickness of few micrometers,[19] which could be due to the fact that the extra laser pulses are required to melt the opaque frost layer.

### 3.3. HSF LIPSSs on silicon

To demonstrate the high-precision processing capability of frost-assisted laser ablation, a square grid with a size of 20×20 μm was inscribed on a silicon plate with tightly focused femtosecond laser at a pulse energy of 1 μJ. The ablated grooves with widths as small as a few microns were produced with a 50× objective lens (N.A. = 0.85) and a single scan with a translation speed of 1 mm/s. At the same time, in order to rapidly make the frost layer melt under low-energy laser irradiation, the thickness of the frost layer was controlled to be as tiny as ~9 μm. **Figure 4**a and 4b show laser ablated microgrooves without frost assistance, and there are a large amount of debris and considerable recast layers around these microgrooves, which cannot be removed even after a 5-min ultrasonic cleaning in water. In contrast, when the silicon wafer was covered with a thin frost layer, both debris deposition and recast layer are fully eliminated, and smooth and uniform microgrooves can be obtained, as shown in Figure 4c. It can be seen from Figure 4d that high-spatial-frequency (HSF) laser-induced periodic surface structures (LIPSSs) with a ripple spacing of 110 ± 10 nm (~λ/9) are uniformly distributed in the whole grooves. To the best of our knowledge, the formation of HSF Si-LiPSSs is previously confined in water [21,22] or at very low laser fluence close to the ablation threshold [23], and it is the first demonstration that large-area uniform HSF Si-LiPSSs at high laser fluence (~25 J/cm$^2$, Supporting Information) could be achieved with assistance of thin frost layer, which offers a potential to achieve high-efficiency and large-area uniform hierarchical nanotexturing.

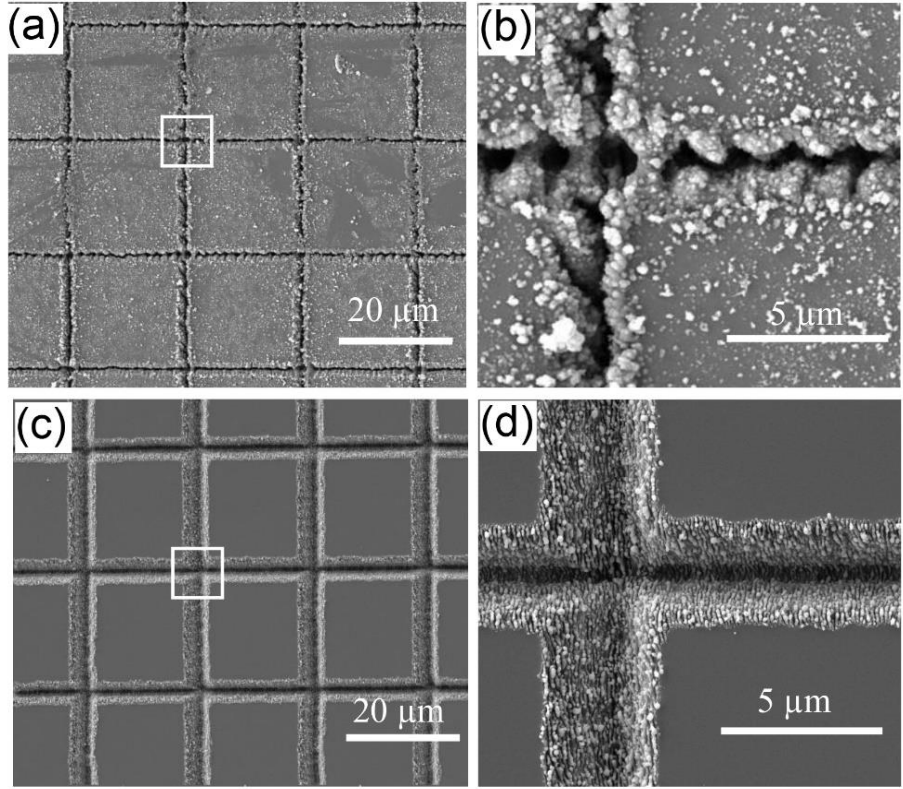

**Figure 4.** SEM micrographs of micro-grids inscribed on silicon covered without (a) and with (c) frost layer. (b) and (d): Zoomed-in images of the area marked by the white rectangle in (a) and (c), respectively.

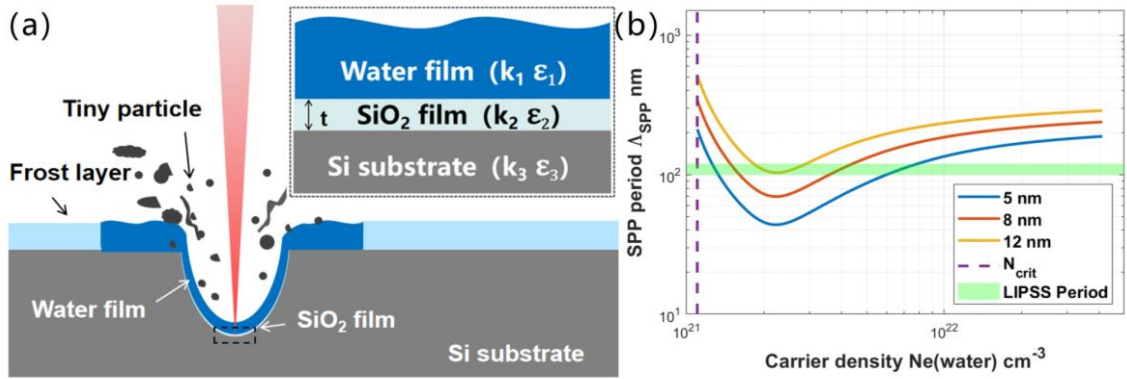

**Figure 5.** (a) Schematic scenario for frost-assisted grooving at high laser fluence, the inset: scheme of the thin-film-SPP model accounting for the HSF LIPPSs. (b) The calculated SPP period as a function of electron density $Ne$ of water film for three different thicknesses of the silicon oxide layer. The green region indicates the experimentally observed LIPSS periods ranging from 100 nm to 120 nm. The purple dashed line corresponds to the critical carrier density threshold of water ($N_{crit}=1.1\times10^{21}$ cm$^{-3}$).

**Figure 5**(a) shows a schematic scenario of frost-assisted femtosecond laser ablation at high fluence. Due to the transient ice-water phase transition induced by femtosecond laser [17], the ablated grooves could be covered by a local water film resulting from melting of peripheral frost layer. Moreover, the ablated silicon surface could be oxidized under high-fluence irradiation, resulting in a thin silicon oxide layer sandwiched between the silicon substrate and the water film, as illustrated in the inset of Figure 5(a). Based on the above scenarios, a thin-film surface plasmon polaritons (SPP) model [24] is used to explain the HSF LIPSS obtained in our experiment. Under high-fluence irradiation of successive laser pulses, the optically excited carrier densities in both silicon substrate and water film could transiently reach their critical values for the SPP excitation, while keeping the thin silicon oxide layer almost transparent. Therefore, it is reasonable to assume the SPPs were excited at both interfaces of the sandwich structure.

An implicit expression of SPP dispersion relation can be derived from the continuity conditions of electromagnetic component as follows [25]:

$$e^{-2k_2 t} = \frac{\frac{k_2}{\varepsilon_2}+\frac{k_1}{\varepsilon_1}}{\frac{k_2}{\varepsilon_2}-\frac{k_1}{\varepsilon_1}} \times \frac{\frac{k_2}{\varepsilon_2}+\frac{k_3}{\varepsilon_3}}{\frac{k_2}{\varepsilon_2}-\frac{k_3}{\varepsilon_3}} \quad (1)$$

where $k_j$ and $\varepsilon_j$ (j=1,2,3) are the complex wave vector component of allowed SPP modes and the complex dielectric permittivity, respectively, corresponding to three different mediums as shown in Figure 5(a), and $t$ is the thickness of the sandwiched silicon oxide layer. The propagation constant β of SPP modes is determined by $k_j^2 = \beta^2 - k_0^2 \varepsilon_j$, and the SPP period $\Lambda_{spp}$ can be calculated by $\Lambda_{spp} = 2\pi/Re(\beta)$. Using a Drude model, the complex dielectric permittivity of the excited silicon and water as a function of carrier density ($N_e$) can be derived as follows [26]:

$$\varepsilon_{j(Ne)} = \varepsilon_{\infty,j} - \frac{N_e e^2}{\varepsilon_0 m^* m} \frac{1}{(\omega^2 + i\omega/\tau)} \quad (2)$$

in which ω is the incident light frequency in vacuum, τ is the Drude damping time of free electrons, and $\varepsilon_0$, $m$, $m^*$ and $e$ are the dielectric permittivity of the vacuum, the electron mass, the optical effective mass of carriers and the electron charge, respectively. $\varepsilon_{\infty,j}$ is the dielectric constant of the non-excited material at the irradiation wavelength written as $\varepsilon_{\infty,j} = (n_j + i\kappa_j)^2$, where $n_j$ and $\kappa_j$ are the refractive index and the extinction coefficient, respectively.

In the numerical calculation of the SPP period, for simplicity, the carrier density of the laser-excited silicon is fixed to $6 \times 10^{21}$ cm$^{-3}$ [24]. The Drude model parameters and the optical constants at the irradiation wavelength of 1030 nm were given as follows: $n_1$=1.321, $n_2$=1.450, $n_3$=3.563, $\kappa_1=\kappa_2$=0, $\kappa_3$=2.2×10$^{-4}$, $m^*_{silicon}$=0.18, $m^*_{water}$=0.5, $\tau_{water}$=1.1 fs, $\tau_{silicon}$=1.7 fs [24, 27]. Figure 6(b) shows the SPP period calculated as a function of the carrier density of the laser-excited water for different thicknesses of the silicon oxide layer. One can see that, at a carrier density of ~2.2×10$^{21}$ cm$^{-3}$, the SPP periods for different thickness of 5 nm, 8 nm and 12 nm simultaneously reach minimum values of ~ 44 nm, 70 nm, and 104 nm, respectively. The results shown in Figure 6(b) suggest that the thickness of sandwiched silicon oxide has a significant influence on the SPP period. When the silicon oxide thickness is below 12 nm, the experimentally observed LIPSS periods of ~110 nm can be achieved at both low and high carrier density, which are in good agreement with our experimental results on HSF LIPSSs at high fluence, also consistent with the previously reported HSF LIPSSs at low fluence [24].

### 3.3. Microtexturing on curved surfaces

The homogeneous frost layer could be formed on the whole cold surfaces exposed to humid air, if the cold surfaces are maintained at uniform temperature below to the freezing point. As a result, the frost-assisted strategy is also applicable to curved surfaces. To demonstrate this capability, a proof-of-principle experiment was performed for microtexturing on cylindrical surface of a borosilicate glass rod by frost-assisted laser ablation. As illustrated in **Figure** 6a, a borosilicate glass rod was clamped in a rotating fixture and was close to a set of V-shape thermoelectric cooler, and a homogenous frost layer was produced by slowly rotating the cooled rod. The high-quality hexagonal micropatterns, as shown in Figure 6b, were inscribed on the cylindrical surface by the combined motion of a rotary motor and a galvo scanner. The femtosecond laser with a pulse energy of 25 μJ was focused by an f-theta lens with a focal length of 100 mm. The ablated grooves with a width of ~12 μm were fabricated with a single scan at a speed of ~100 mm/s. Figure 6c and 6d present zoom-in micropatterns fabricated by traditional and frost-assisted micro-texturing, respectively. It can be clearly observed that there are debris redeposition zones with a width of ~ 30 μm along the grooves, while the textured surface with frost assistance is quite clean and debris-free. The high-quality and high-efficiency surface microtexturing on curved surfaces will find many practical applications in anti-friction bearings and wear-resistant tools.

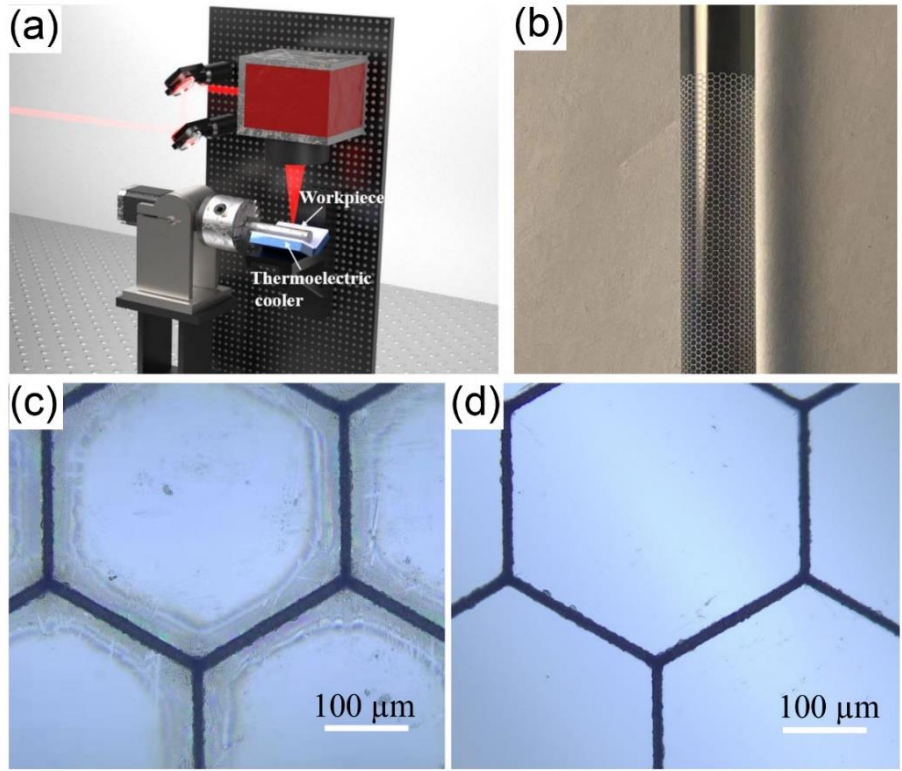

**Figure 6.** (a) Schematic of frost-assisted femtosecond laser ablation on cylindrical surfaces; (b) A borosilicate glass rod textured with hexagonal micropatterns by frost-assisted femtosecond laser ablation; Optical microscopic images of the micropattern without (c) and with (d) the assistance of frost.

## 4. Conclusion

In this work, we present a novel frost-assistance strategy to tackle the thermal accumulation issues for femtosecond laser ablation at high laser fluence, such as redeposition of ablated debris and formation of recast layer. A thin frost layer was preliminarily formed by condensation of ambient water vapor onto the sample surfaces with a temperature below the freezing point. When femtosecond laser is focused onto the frost-covered surfaces, the local frost layer around the laser spot will melt into water film, helping to boost ablation efficiency, suppress recast layer and reduce debris redeposition. The experiment results demonstrate that the frost-assistance strategy enables high-quality and high-efficiency surface texturing at high laser fluence, and also is applicable to curved surfaces and various different materials (Supporting Information). It is noteworthy that the redeposition of debris and the formation of recast layer can also be improved by lowering the laser fluences to the ablation threshold or decreasing the repetition rate of laser pulses to suppress thermal effect, but these methods will significantly reduce the efficiency of surface texturing.

With the assistance of a thin frost layer, HSF LIPSSs on silicon can be induced by a single fast scan with high laser fluence, which offers a potential to achieve high-efficiency and large-area uniform hierarchical nanotexturing. A thin-film SPP model is successfully used to explain the HSF LIPSSs obtained in our experiment, suggesting that the thickness of sandwiched silicon oxide layer has a significant influence on the SPP period. Noting that the intrinsic frost formation process makes its thickness controllable and homogeneous at large scales, we speculate that the frost-assisted femtosecond laser ablation could open a new way to large-area uniform and high-quality 2D and 3D surface micro/nano-structuring across a broad range of applications.

**Acknowledgements**

The authors would like to acknowledge funding support from National Science Foundation of China (NSFC) under Grant No. 61675220. Wenhai Gao and Kai Zheng contributed equally to this work.

**Conflict of Interest**

The authors declare no conflicts of interest.

Supporting Information

# High-quality femtosecond laser surface micro/nano-structuring assisted by a thin frost layer

*Wenhai Gao[#], Kai Zheng[#], Yang Liao\*, Henglei Du, Chengpu Liu, Chengrun Ye, Ke Liu, Shaoming Xie, Cong Chen, Junchi Chen, Yujie Peng, and Yuxin Leng\**

## 1. Thickness comparison of frost and ice layer

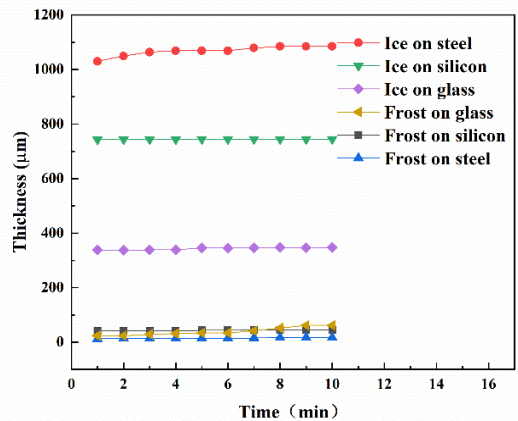

**Figure S1.** Thickness of frost and ice layers on the surface of different materials as a function of cooling time.

The growth of frost and ice layers on the surfaces of different materials were experimentally investigated (Figure S1). In this experiment, the ice layers were formed as follows: different materials were soaked in pure water and then raised from the water at a certain speed, and finally placed on a semiconductor refrigeration sheet. After the same cooling time, the thicknesses of the ice and frost layers were measured using a three-axis measuring microscope with a resolution of 1 µm. The relative humidity and temperature of ambient were kept constant at 35% and 25°C. By controlling the surface temperature at -5°C, the thickness of frost layer can be kept at 10-20 µm and can roughly constant over 10 minutes, while the thickness of ice layer ranges from 300-1100 µm. We expect the thickness of ice layer is related to the surface wetting properties of different material.

The slow rate of frost growth is also a prerequisite for stable laser processing. According to M. Kandula's model [1], the frost growth rate was derived as follows:

$$\frac{dx_s}{dt} = \frac{h_m(\omega_a - \omega_s)}{\rho_f \left(1 + x_s \left[c_2(1-\xi^{0.5}) + \frac{1}{\theta}\right] \frac{1}{(T_m - T_w)} \frac{dT_{fs}}{dx_s}\right)}$$ (S1)

Where, $x_s$ is the frost thickness, $\rho_f$ is the frost density, $h_m$ is the convective mass transfer coefficient, $\omega_a$ and $\omega_s$ are the humidity ratios in the atmospheric air and at the frosted surface respectively; $\theta = \frac{T_{fs} - T_w}{T_m - T_w}$ is the dimensionless frost surface temperature, $Tm$ is the melting temperature, $T_w$ is the temperature at the cold surface, $T_{fs}$ is the frost surface temperature; $\xi = \frac{R_e}{R_{e_{cr}}}$, $Re$ is Reynolds number of air flow, $Re_{cr}$ is the critical Reynolds number for laminar turbulent transition. The equation of the frost growth rate indicates that the slow rate of frost growth depends on high frost surface temperature, low ambient humidity, and low convective mass transfer coefficient.

## 2. Estimation of laser fluences

To achieve Si-LIPSSs as shown in Figure 4e in the manuscript, the laser power was set at 100 mW which is equivalent to a pulse energy (E) of 1 µJ/pulse at a 100 kHz repetition rate. The spot sizes (2ω$_0$) for laser ablation on silicon was measured to be 3.2 µm. According to the equation of laser fluence $F = 2E/\pi \cdot w_0^2$ [2], a laser fluence of 25 J/cm$^2$ was derived, much higher than the ablation threshold of Si (0.28±0.3 J/cm$^2$ [3]).

## 3. Frost-assisted femtosecond laser ablation of fused silica glass

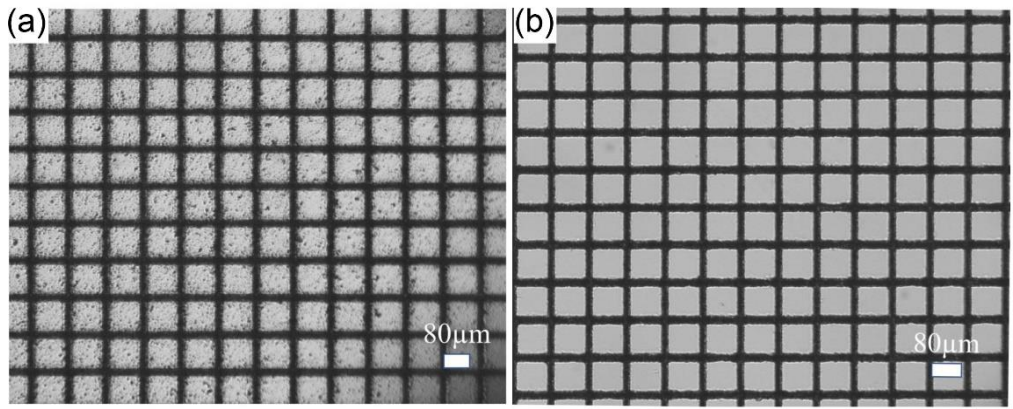

**Figure S2.** Optical microscope images of microgrids in fused silica glass by femtosecond laser ablation (a) in air and (b) under frost.

Figure S2 presents a comparison of the microgrids etched in the glass without and with frost assistance. The femtosecond pulsed laser was set at a pulse energy of 5 μJ and a repetition rate of 100 kHz. The ablated grooves were produced with a 5× objective lens (N.A. = 0.1) and a single scan with a translation speed of 1 mm/s. The heat-effected zone and debris redeposition were noticeably improved under frost compared with the case in air.

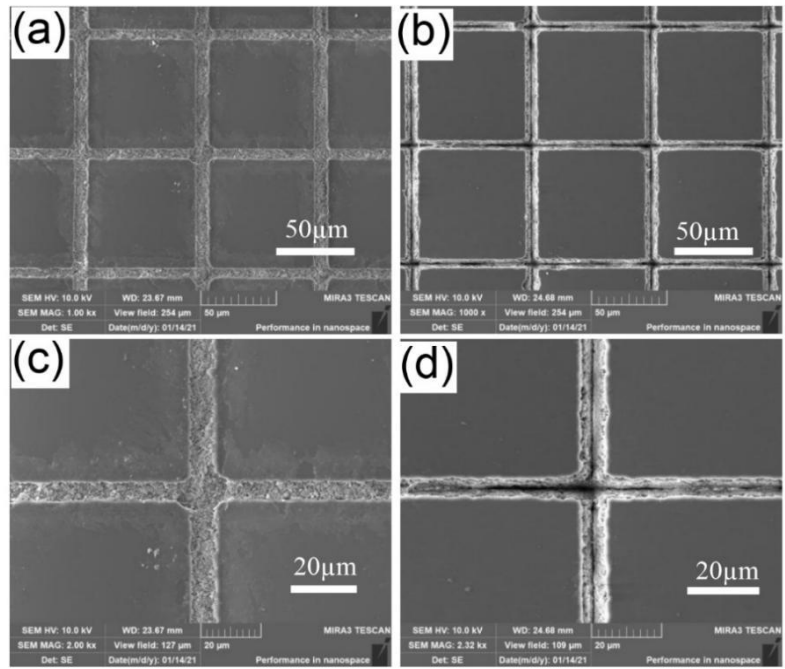

**Figure S3.** SEM micrographs of micro-grids etched in fused silica glass without (a), (c), and with (b), (d) frost layer.

Figure S3 presents SEM images of micro-grids in fused silica glass inscribed in air and under frost with the same ablation parameters as in Figure S2. It can be clearly observed that the microgrids with the assistance of frost have clear edges, and there is no any debris in the ablated grooves, showing higher machining quality and efficiency compared to laser etching in air.

## 4. Frost-assisted femtosecond laser drilling on an aluminum foil

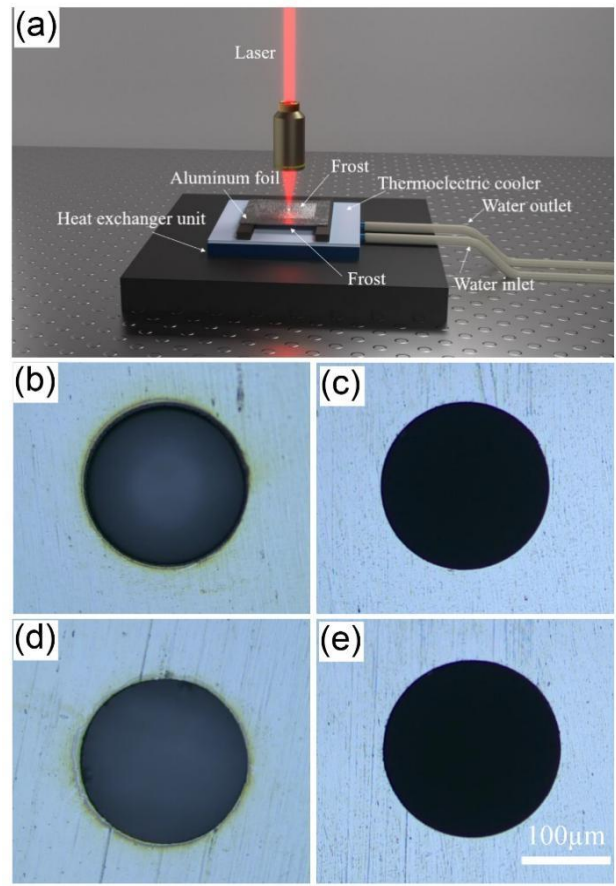

**Figure S4.** (a) Schematic of frost-assisted laser through-hole drilling on a thin aluminum foil; optical microscopic images of entrance surface of a through-hole drilled (b) in air (c) under frost; optical microscopic image of the exit surface of a through-hole drilled (d) in air (e) under frost.

Another advantage of frost-assisted femtosecond laser ablation is its suitability for through-hole drilling, because thin frost layer can be formed on both entrance and exit surfaces of the drilled materials, on condition that these surfaces are below the freezing point and exposed in air. Figure S4(a) presents an experimental setup for frost-assisted laser through-hole drilling on a thin aluminum foil with a thickness of 10 µm. To make both sides of the aluminum foil exposed in air, the foil was suspended by two supports fixed on a thermoelectric cooler, and thin frost layers with the same thickness of ~10 µm were simultaneously formed on both sides. Figure S4(b-e) present comparison results on the entrance and exit surfaces of a through-hole drilled without and with frost assistance. It can be seen that there are obvious heat-affected zones and burrs on both the entrance and exit surfaces for the through-hole drilled in air, while the through-hole drilled under frost has sharp or rounded edges on its both surfaces.